\documentclass[amssymb,one column]{revtex4}                  
\usepackage{graphicx}
\usepackage{amsmath}
\usepackage{float}
\usepackage{subfigure}
\usepackage{graphicx,epstopdf}

\newcommand{\qlt}{\overline{q_t(L)}}
\newcommand{\qmean}{\overline{q_t}}

\newcommand{\qfinal}{\overline{q_\infty(L)}}

\newcommand{\cm}{\theta_{h}}

\newcommand{\qfinalN}{\overline{q_\infty(N)}}
\newcommand{\qfinalNmax}{\overline{q_\infty(N_{\rm max})}}
\newcommand{\qfinalinfN}{\overline{q_\infty(N\rightarrow\infty)}}

\begin{document}

\title{Nature vs.~Nurture in Discrete Spin Dynamics}

\author{D.L.~Stein} 
\email{daniel.stein@nyu.edu}
\affiliation{Department of Physics and Courant Institute of Mathematical Sciences,
New York University, New York, NY 10012 USA and NYU-ECNU Institutes of Physics and Mathematical Sciences at NYU Shanghai, 3663 Zhongshan Road North, Shanghai, 200062, China}

\begin{abstract}
The problem of predictability, or ``nature vs.~nurture'', in both ordered and disordered Ising systems following a deep quench from infinite to zero temperature is reviewed. Two questions are addressed. The first deals with the nature of the final state: for an infinite system, does every spin flip infinitely often, or does every spin flip only finitely many times, or do some spins flip infinitely often and others finitely often? Once this question is determined, the evolution of the system from its initial state can be studied with attention to the issue of how much information contained in the final state depends on that contained in the initial state, and how much depends on the detailed history of the system. This problem has been addressed both analytically and numerically in several papers, and their main methods, results, and conclusions will be reviewed. The discussion closes with some open problems that remain to be addressed.
\end{abstract}

\maketitle

\section{Introduction}
\label{sec:intro}

It is a great pleasure to contribute to this volume in honor of Chuck Newman's $70^{\rm th}$ birthday. Chuck has made so many fundamental contributions to 
probability theory, mathematical statistical mechanics, percolation theory, and many related fields that it was difficult to choose which topic to write about. The
problem was resolved by settling on a problem of longstanding interest to Chuck, and one in which he remains heavily involved: the nonequilibrium dynamics of
interacting spin systems, in particular following a {\it deep quench\/}, in which a system at high temperature is rapidly cooled to low temperature, after which it 
evolves according to equilibrium dynamics.  

Our initial foray into this problem appeared in~\cite{NS99a}, although we had earlier looked at the closely related problems of persistence~\cite{NS99b} and 
(somewhat later) aging~\cite{FINS01}.  Our initial interest was in determining how and under what conditions equilibration occurs (or doesn't occur) in an infinite system following a deep quench.
One has to specify first exactly what one means by ``equilibration in an infinite system''. Our proposal was that one can sensibly talk about a thermodynamic system settling down to
an equilibrium state in the sense of {\it local equilibration\/}: that is, fix a region of diameter $L$ and ask whether after a finite time $\tau(L)$  domain walls cease to sweep across the region.
If the answer is yes, i.e., if every finite $L$ corresponds to a $\tau(L)<\infty$ after which no spins inside the region ever flip again, then the infinite system locally equilibrates.
(It's not only perfectly fine, but also expected in most cases, that $\tau(L)\to\infty$ as $L\to\infty$.)

The main thrust of~\cite{NS99a} then went off in a different direction, but it set the seeds for further examination of the problem of local equilibration, its presence or absence in given systems, and its consequences.
A paper with Seema~Nanda soon followed that focused strictly on zero temperature~\cite{NNS00} (\cite{NS99a} studied positive temperature only), with numerous results that began the sorting of systems in which 
local equilibration occurred and those in which it did not. At zero temperature, a new set of issues and problems arises, in particular the possibility of trapping in a metastable state~\cite{NS99c}; nevertheless,
if every spin flips only a finite number of times, then local equilibration has occurred, regardless of whether the final state is metastable or a global minimum. 

\section{Local equilibration, weak nonequilibration, and chaotic time dependence}
\label{sec:lne}

The first question to be resolved is then whether local equilibration occurs for a given system. When it does not occur, then there are further questions to be resolved.
There are two possibilities when local {\it non\/}equilibration (LNE) manifests itself:  if one averages over all dynamical realizations, does the 
{\it dynamically averaged\/} configuration have a limit --- i.e., is there a limiting distribution of configurations? 
Or does even the distribution not settle down?  We refer to the first possibility as ``weak LNE'', while the second
is referred to as {\it chaotic time dependence\/} (CTD)~\cite{NS99a}. Weak~LNE implies a complete lack of predictability, 
while CTD implies that some amount of predictability remains~\cite{NS99a,NNS00}.

These claims are justified as follows.  Consider for specificity the infinite uniform ferromagnet, in which there
are only two dynamical final states: all spins up or all spins down. If weak LNE
occurs, then (because of global spin-flip symmetry) after some time in a fixed finite region roughly half the dynamical
realizations are in the plus state and half are in the minus state, and
this remains true for all subsequent times (though in any particular
dynamical realization, the system never settles down into either).

On the other hand, if CTD occurs, then the dynamical averaging fails
to fully mix the two possible outcomes at any time; a time-independent distribution is never achieved. In that case,
given the initial configuration, one retains some predictive power at
arbitrarily large times. These will presumably reflect fluctuations
favoring one phase or the other (in that particular region) in the
initial state. Compare this to weak LNE, which requires stronger mixing
that destroys information contained in these initial
fluctuations. Because at zero temperature the configuration evolves
according to a deterministic set of rules (given the initial
configuration and a realization of the dynamics), the lack of a limit
of the averaged configuration in CTD corresponds to the usual notions
of deterministic chaos, hence the name ``chaotic time dependence''. It
is amusing that in this context, chaotic time dependence implies {\it
  greater\/} predictability than its nonchaotic alternative.

These considerations motivate the following specific question, first proposed
in~\cite{NS99a}: given a typical initial configuration, which then
evolves under a specified dynamics, how much can one predict about the
state of the system at later times?  We have colloquially referred to
this as a ``nature~vs.~nurture'' problem, with ``nature"
representing the influence of the initial configuration and ``nurture''
representing the influence of the random dynamics.

The study of nature vs.~nurture therefore 
provides interesting information on some central dynamical issues
concerning different classes of models.  This problem is related to the
general area of phase ordering kinetics~\cite{Bray94}.  In particular, Krapivsky, Redner and
collaborators~\cite{SKR00,SKR01} investigated both the $2D$ and $3D$~Ising
models with zero temperature Glauber dynamics to understand the time scales
and final states of the dynamics.  Derrida, Bray and Godr\`{e}che introduced
the {\em persistence exponent}~\cite{DBG94}, which characterizes the power
law decay of the fraction of spins that are unchanged from their initial
value as a function of time after a quench.  This exponent was measured for
the zero temperature $2D$~Ising model by Stauffer~\cite{Stauffer94} and
calculated exactly for $1D$ by Derrida, Hakim and Pasquier~\cite{DHP95}.

\section{Uniform vs.~disordered systems}
\label{sec:uniform}

From here on, we restrict our attention to Ising systems at zero temperature and in the absence of any external field, where the bulk (but not all) of our studies have focused.
In every case, we start with an infinite temperature spin configuration; i.e., the starting configuration is a realization of a Bernoulli process in which each spin is chosen independently of 
the others following the flip of a fair coin. The ensuing evolution of the system is governed by zero-temperature Glauber dynamics, in which each spin has an attached Poisson clock with rate~1. 
When a spin's clock rings, it looks at its neighbors and computes the energy change $\Delta E$ associated with flipping (with all other spins remaining fixed).
If the energy decreases ($\Delta E< 0$) as a result of the flip, the flip is carried out. 
If the energy increases ($\Delta E> 0$), the flip is not accepted. If the energy remains the same ($\Delta E = 0$), 
a flip is carried out with probability~1/2. This last ``tie-breaking'' rule can occur only for models (such as the homogenous ferromagnet, or the $\pm J$ spin glass) 
in which a zero-energy flip can occur; for models with continuous coupling distributions, there is zero probability of a tie. (This is also the case for uniform systems 
with an odd number of neighbors, such as the $2D$ uniform ferromagnet on the honeycomb lattice.) 

For a uniform system, there are two sources of randomness in the problem: the choice of realization of the initial spin configuration, and the realization of the dynamics (i.e., the order
in which Poisson clocks for different spins ring, and for models where zero-energy flips are possible, the results of the tie-breaking coin flips). For disordered systems, a third source of randomness enters 
in the choice of realization of the couplings.

As indicated in the Introduction, the first question that must be resolved is whether the system under study equilibrates locally.  
This is generally a difficult problem, and our knowledge remains incomplete. Nevertheless, we have obtained results on several
systems that include some of the most studied in statistical mechanics.

The simplest is the $1D$ Ising chain. For the uniform ferromagnetic chain, it was proved in~\cite{NNS00} (although the result was known earlier~\cite{Arratia83,CG86}) that there is no local equilibration: every spin flips infinitely often.
In general dimension, the Hamiltonian is
\begin{equation}
\label{eq:ferro}
{\cal H}=-\sum_{\langle x,y\rangle}\sigma_x\sigma_y\, ,
\end{equation}
where $\sigma_x=\pm 1$ is the spin at site $x$ and $\langle\cdot\rangle$ denotes a sum over nearest neighbor spins only.

The proof that every spin flips infinitely often is straightforward and will be informally summarized here. (For the remainder of the paper, most results will be quoted, with references to the original papers for
the detailed proofs.) To begin, it is easy to see that the only fixed points of the dynamics are $\sigma_x=+1$ for all~$x$ or $\sigma_x=-1$ for all~$x$. Let $\omega$ denote a realization of the dynamics and $\sigma^0$ a realization of the initial conditions. Then, because both are i.i.d., their joint distribution~$P_{\sigma^0,\omega}$ is translation-invariant and translation-ergodic. Because the events that the system lands in the all-plus or the all-minus final states are each translation-invariant, and because both the initial condition and the dynamics are invariant under a global spin flip, each of these events has $P_{\sigma^0,\omega}$-probability zero.

Now let $A_x^+$ be the event that $\sigma_x$ flips only finitely many times and its final state is $+1$; similarly for $A_x^-$.  By the same reasoning as above, we must have  $P_{\sigma^0,\omega}(A_x^+)=P_{\sigma^0,\omega}(A_x^-)=p$, with $0\le p\le 1/2$. If $p>0$, then there must exist a pair of spins at two distinct sites $x$ and $x'$ with opposite final states. This implies a domain wall somewhere between the two that never moves past either one. But it is easy to see that, with positive probability in the dynamics, a sequence of clock rings exists that moves the domain wall outside of this confined range, leading to a contradiction.

We will call systems in which every spin flips infinitely often as being in class ${\cal I}$; if every spin flips only finitely often, we refer to it as being in class ${\cal F}$. There are also systems in which some spins flip infinitely often and others only finitely often; we refer to these as being in class~${\cal M}$. In~\cite{NNS00} it was proved that the $2D$ uniform ferromagnet also belongs to class~${\cal I}$; the proof is similar in spirit to the $1D$ case but is slightly more elaborate.

What about the uniform ferromagnet in higher than two dimensions? This remains an open problem. Rather old numerical studies~\cite{Stauffer94} indicate that above five dimensions these models are no longer in class~${\cal I}$, but the situation remains unclear. The possible presence of dynamical fixed points with complicated geometries in three dimensions and above have so far prevented further analytical progress.

Before leaving the subject of the uniform ferromagnet, it's worth noting that Chuck and collaborators have looked at other mathematically interesting situations, including quasi-$2D$ slabs of varying thicknesses and boundary conditions (which they showed are either type~${\cal F}$ or ${\cal M}$ depending on thickness and boundary condition)~\cite{DKNS13}, and on $\mathbb{Z}^2$ with a single fixed spin (class~${\cal I}$, modulo the frozen spin).

\bigskip

Our discussion has focused so far on uniform ferromagnets. What about disordered systems, in particular, random-bond Ising models? We consider two important classes: random ferromagnets, in which the couplings are i.i.d.~nonnegative random variables, and spin glasses, where couplings can be positive or negative. (Minor point: in the latter case, if the coupling distribution is asymmetrically distributed about zero, one will have a ferromagnet if the average ratio of positive to negative couplings is sufficiently high. But it turns out this will have no effect on determining whether equilibration occurs.) The Hamiltonian is now
\begin{equation}
\label{eq:disordered}
{\cal H}=-\sum_{\langle x,y\rangle}J_{xy}\sigma_x\sigma_y\, ,
\end{equation}
where the $J_{xy}$ are independent random variables chosen from a common probability distribution.

If one is interested in thermodynamic behavior --- e.g., presence or absence of a phase transition, or the number of pure states at a fixed positive temperature and dimension --- then central limit theorem-type considerations lead one to expect that the form of the coupling distribution is unimportant, as long as certain features of the distribution, such as mean and variance, are unchanged. So, for example, one expects the same thermodynamic behavior when the coupling distribution is the normal distribution~$N(0,1)$ or $\pm 1$, the latter referring to a distribution where each coupling is assigned the value $\pm 1$ independently with the flip of a fair coin.

However, from the point of view of dynamics, particularly with regard to the question of whether local equilibration occurs, the differences between distributions becomes important. In particular, Theorem~3 in~\cite{NNS00} shows that, under very mild conditions (existence of a finite mean), {\it any\/} model with Hamiltonian~(\ref{eq:disordered}) and with i.i.d.~couplings chosen from a continuous distribution will belong to class~${\cal F}$.

In fact the proof is more general, showing that in any discrete-spin model there are only finitely many flips of any spin resulting in a {\it nonzero\/} energy change. This immediately implies that models belonging to class ${\cal I}$ (or ${\cal M}$), such as the $1D$ and $2D$ uniform ferromagnets, do so exclusively because of an infinite number of zero-energy flips at each site (or a subset of sites). Consequently, models with continuous coupling distributions all belong to class~${\cal F}$, because the chance of a tie at any site has zero probability; but the same conclusion applies, say, to the above-mentioned $2D$ uniform ferromagnet on a honeycomb lattice, where the chance of a tie is also zero.

Consequently, the question of the ultimate dynamical fate of a disordered system with continuous coupling distribution is answered in all dimensions.

The restriction to finite mean was needed for technical reasons in the proof, but may not be necessary. Also in~\cite{NNS00} it was proved that the so-called {\it highly disordered\/} models~\cite{NS94,BCM94,NS96} in any dimension belong to class~${\cal F}$. These are models in which the coupling magnitudes are sufficiently ``stretched out'' so that the magnitude of any coupling is at least twice the value of the next smaller magnitude and no more than half that of the next larger magnitude. (For a formal definition, see~\cite{NNS00}.) Of course, this needs to be done in a size-dependent manner, rescaling the values of the couplings as one increases the volume under consideration~\cite{NS94,NS96}. It turns out that $1D$ chains with continuous coupling distributions, although not satisfying the above criterion, still belong to this class. That is, $1D$ chains with {\it any\/} continuous coupling distribution, regardless of whether the mean is finite, will fall into class~${\cal F}$.

As a final remark on continuous coupling distributions in general, an important point is that it doesn't matter (for the purposes of whether all spins eventually fixate) whether one is talking about a random ferromagnet or a spin glass --- the signs of the coupling do not enter into these considerations, only the coupling magnitudes.

We close this section by mentioning a result for a model with discrete disorder: the $\pm J$~spin glass in $2D$. It was shown in~\cite{GNS00} that this model belongs to class~${\cal M}$. We are unaware of any other results on this class of models in terms of their final evolutionary states.

\section{Nature vs.~nurture}
\label{sec:nvsn}

Given the classification of systems in the previous section, we turn now to the main focus of this review --- namely, the extent to which information 
contained in a configuration at time~$t$ can be inferred from knowledge of the initial state of the system. We first need to quantify this concept,
and the approach used rests on the idea of determining what proportion of the information contained in the state at time $t$ depends on the initial
condition and what proportion on the realization of the dynamics.

If the system is type~${\cal I}$, one might expect that the information contained in the initial state will decay to zero as $t\to\infty$. We will discuss below how
this is indeed the case in the $1D$ and $2D$~uniform ferromagnets, where the decay takes the form of a power law. This leads to a
new exponent characterizing the power-law decay, which we have denoted the {\it heritability exponent\/}. We will return to the heritability exponent in Sect.~\ref{subsec:heritability}.

We turn next to the (perhaps) simpler case of type-${\cal F}$ systems.
Because these by definition equilibrate locally, we compare the final state 
with the initial state, averaged over
many dynamical trials, to determine to what degree initial information has
been retained. In~\cite{NNS00} we introduced a type of dynamical order
parameter, denoted $q_D$ (here ``$D$'' refers to dynamical, not dimension).

Let $\sigma^0$ denote the realization of the initial condition and $\sigma^t$ the configuration at time
$t$ later. This of course depends on the dynamical realization $\omega$, but to keep the notation 
from becoming too unwieldy, we suppress the dependence of
$\sigma^t$ on $\sigma^0$ and $\omega$. Let $\langle\sigma_x\rangle_t$
denote the state of $\sigma_x$ averaged over all dynamical realizations up to time $t$ for a
{\it fixed\/} $\sigma^0$. We then study the resulting
quantity averaged over all initial configurations and (if the system is disordered) coupling
realizations. 

Denoting the latter averages (with respect to the joint
distribution $P_{{\cal J},\sigma^0} = P_{\cal J} \times P_{\sigma^0}$)
by ${\bf E}_{{\cal J},\sigma^0}$, we define $q_D = \lim_{t \to
  \infty}q_t$ (providing the limit exists), where
\begin{equation}
\label{eq:qt}
q_t = \lim_{L \to \infty} |\Lambda_L|^{-1}\sum_{x \in \Lambda_L}
(\langle \sigma_x \rangle_t)^2 = {\bf E}_{{\cal J},\sigma^0}
(\langle \sigma_x \rangle_t^{\,2})
\end{equation}
and $\Lambda_L$ is a $d$-dimensional cube of side $L$ centered at the
origin.  The equivalence of the two formulas for $q_t$ in~(\ref{eq:qt}) follows from
translation-ergodicity~\cite{NNS00}.

The order parameter $q_D$ measures the extent to which $\sigma^\infty$ is determined by $\sigma^0$ rather than by $\omega$; it is a dynamical analog to the usual EdwardsÐAnderson order parameter. 

\subsection{$1D$ and highly disordered models}
\label{subsec:1D}

The dynamical order parameter can be exactly calculated for the $1D$~Ising chain. For the uniform ferromagnet, $\sigma_x^\infty$ does not exist. But it does when the couplings are drawn from a continuous distribution 
(the details of the distribution are unimportant, as long as its support is continuous), and moreover for these models $q_D=1/2$. The result and proof appear in~\cite{NNS00}, but the argument underlying the result is easy to state informally.
The basic idea is to note that every spin lives in the domain of influence of a ``bully bond''. This is a coupling whose magnitude is larger than the couplings to either side; such a coupling must be satisfied in any fixed point of the 1-spin-flip
dynamics (and more generally, in any ground state). As one moves along the chain starting from either side of the bully bond, couplings decrease in magnitude until arriving at a local minimum: i.e., a coupling whose magnitude is {\it smaller\/} than the couplings to either side. (Of course, the bond whose coupling value is a local minimum could be adjacent to the bully bond.)

The domain of influence of the bully bond is then the set of spins that live on the sites of the bonds to either of its sides until one reaches the nearest local minimum bonds to its right and left. For the local minimum bond to the right of the bully bond, the spin on its lefthand site belongs to the domain of influence of the bully bond; for the local minimum bond to the left of the bully bond, the spin on its righthand site belongs to the domain of influence of the bully bond. (The remaining two spins on the two local minimum bonds belong to other domains of influence.) In this way the entire chain is partitioned into domains of influence, each ``governed'' by a single bully bond.

Regardless of the dynamical realization, it must be the case that once the bully bond is satisfied, the final state of every spin in its domain of influence is completely determined. Now consider the initial state $\sigma^0$. In a.e.~realization
of the initial state, half the bully bonds will be satisfied and half will be unsatisfied. For those that are satisfied, the final state of every spin in its domain of influence is completely determined by the initial state; i.e., a.e.~realization of the dynamics will result in the same final state. For those that are unsatisfied, the final state is determined by which Poisson clock, of the two spins on the sites connected to the bully bond, rings first. Since these two events have equal probability, and since the two possible outcomes give equal and opposite contributions to the twin overlap, all spins in such domains of influence contribute zero to $q_D$. Consequently, $q_D=1/2$ for the continuously disordered $1D$~chain: in $\sigma^\infty$ half the final states of the spins are completely determined by the initial configuration, and half are completely determined by the dynamics.

For highly disordered models in any dimension, $q_D$ is also 1/2. The main idea behind the proof is similar to that above, but more involved. Here, one can define {\it influence clusters\/} of similarly defined bully bonds, but one needs to show as well that these influence clusters do not percolate. Details can be found in~\cite{NNS00}.

Before leaving this section, it is worth noting that in models with continuous disorder on the Euclidean lattice $\mathbb{Z}^d$ with $1<d<\infty$, an argument similarly based on bully bonds can be used to show that $0<q_D<1$. In any dimension, the density of bully bonds is strictly positive (though decreasing as dimension increases). If $\rho_b(d)>0$ is the density of bully bonds in $d$~dimensions, then considerations similar to those above provide a lower bound of $(1/2)\rho_b(d)$ and an upper bound of $1-(1/2)\rho_b(d)$ for $q_D$. 

Needless to say, these are poor upper and lower bounds; the main point is that $0<q_D<1$ for disordered models in any finite dimension. We return to this discussion is Sect.~\ref{sec:disordered}.

\subsection{Heritability, damage spreading, and persistence}
\label{subsec:heritability}

We turn now to type-${\cal I}$ systems, in which $\sigma_x^\infty$ does not exist
for any $x$, and therefore $q_D$ is not defined. However, $q_t$ remains perfectly
well-defined for all finite $t$, and one can study this quantity to determine how the initial information
contained in $\sigma^0$ changes with time. 

If one is studying the system numerically (necessary in most cases), then one can model $q_t$ in the following way: 
prepare two Ising systems with
the same initial configuration, and then allow them to evolve
independently using zero-temperature Glauber dynamics. One then computes the spin
overlap between these ``twin'' copies, chooses another initial condition, and so on, eventually 
computing a twin overlap over many different initial conditions.  This overlap as a function of time, which we refer to as the
``heritability'', is essentially the same as $q_t$, and for the $2D$~uniform ferromagnet was found to decay as a power
law in time~\cite{YMNS13}. We denote the exponent
$\theta_h$ associated with the power-law decay of heritability as the
``heritability exponent''.

Heritability is the opposite of ``damage
spreading''~\cite{Creutz86,SSKH87,Grassberger95}, given that the latter involves
starting with two slightly different initial configurations and letting
them evolve with the {\it same\/} dynamical realization. In damage spreading studies one is interested in the extent of the
spread of the initial difference throughout the system as time proceeds.

The nature vs.~nurture question is also related to {\it persistence\/}~\cite{DBG94}, which studies the fraction of spins that 
have not flipped up to time $t$. This was found to decay as
a power law in a number of systems, in particular uniform ferromagnets and
Potts models in low dimensions, and the associated decay exponent
$\theta_p$ is known as the ``persistence exponent''. Although the concepts are related, heritability is not the same as
persistence, which asks which spins have not flipped up to a
time $t$. In contrast, heritability asks to what extent the
{\it information\/} contained in the initial state persists up to time
$t$. A spin may have flipped multiple times during this interval but its
final state might still be predictable knowing the initial condition. 

\section{The $1D$ and $2D$ uniform ferromagnets}
\label{sec:1D2D}

Both the persistence and heritability exponents can be computed exactly in
the $1D$ uniform Ising ferromagnet. It was shown in~\cite{DHP95,DHP96}
that $\theta_p=3/8$ for this system. Correspondingly, it can be
shown that $\theta_h = 1/2$, as discussed in~\cite{YMNS13}, by using
the mapping to the voter model and coalescing random walks (see,
e.g.,~\cite{DHP96,FINS01}).

The $2D$~uniform Ising ferromagnet on the square lattice 
is considerably more difficult, and requires numerical study. 
The (finite-volume) absorbing states include not only the uniform plus and minus states, but also
``striped'' configurations which appear in roughly 1/3 of the runs~\cite{SKR00}. A striped state has one (or more) vertical or
horizontal stripes (but not both) whose boundaries constitute domain
walls separating regions of antiparallel spin orientation.  

An initial study of the nature vs.~nurture problem was reported in~\cite{ONSS06}, in which 
evidence was found for chaotic time dependence. However, the problem was not
fully analyzed and solved until almost a decade later, when Ye~{\it et al.\/}~\cite{YMNS13}
showed that a power-law decay of initial information did occur and computed the
heritability exponent.

Ye~{\it et al.\/}~did twin studies on 21~$L\times L$ squares, from $L=10$ to $L=500$.  For each size 
30,000~runs on independent pairs of twins were taken out to times such that (almost) all
of the samples landed in an absorbing state.  (For each initial condition 
only two dynamical trajectories were computed, one for each twin.)  Because these runs were done on finite lattices, 
the authors studied $q_t(L)= \frac{1}{N}\sum_{i=1}^{N}\sigma_{i}^1(t)\sigma_{i}^2(t)$, where $\sigma_{i}^1(t)$ denotes the state of the $i^{\rm th}$ spin
at time $t$ in twin~1, and similarly for $\sigma_{i}^2(t)$. 

For each pair of twins $q_t(L)$ 
was computed, and averaging over all runs gave the average~$\overline{q_t(L)}$.
Ye~{\it et al.\/}~studied both the size dependence of the final
overlap, $\qfinal = \lim_{t \rightarrow \infty} \overline{q_t(L)}$ and
the time dependence of the infinite volume limit~$\qmean = \lim_{L\rightarrow \infty} \overline{q_t(L)}$.  
It was also shown analytically, and confirmed numerically, that the
behavior of $\qfinal$ and $\qmean$ are connected by a finite size
scaling ansatz~\cite{YMNS13}.

\begin{figure}[h]
\centering
\includegraphics[scale=1]{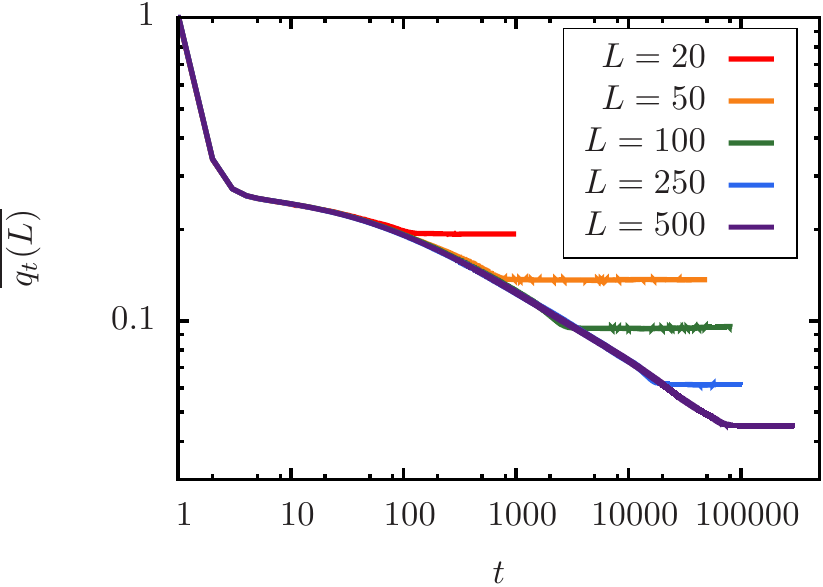}
\caption{$\qlt$\ vs.\ $t$ for several $L$ and for quenches to $T=0$.  The  plateau value decreases from small to large $L$. From~\cite{YMNS13}.}
\label{fig:MTF}
\end{figure}

Consider first the behavior of $\qlt$
vs.~$t$ for several~$L$, shown in~Fig.~\ref{fig:MTF}.
For short and intermediate
times, $\qlt$ appears to follow a single curve for all~$L$, until an
$L$-dependent time scale when $\qlt$ separates from the main curve and
a plateau is reached.  Ye~{\it et al.\/}~made the natural assumption that the single
curve represented the infinite volume behavior~$\qmean$ to good
approximation. The long-time behavior of~$\qmean$ is well-described by a power law of the form $\qmean=d
t^{-\cm}$, with $d = 0.62(3)$ and $\cm= 0.225(6)$ computed at the largest size studied ($L=500$).  The error bars
are obtained by the bootstrap method.  Using other large sizes also, the heritability exponent describing the decay of the
overlap with time was found to be $\cm=0.22\pm 0.02$.

\begin{figure}[h]
\centering
\includegraphics[scale=1]{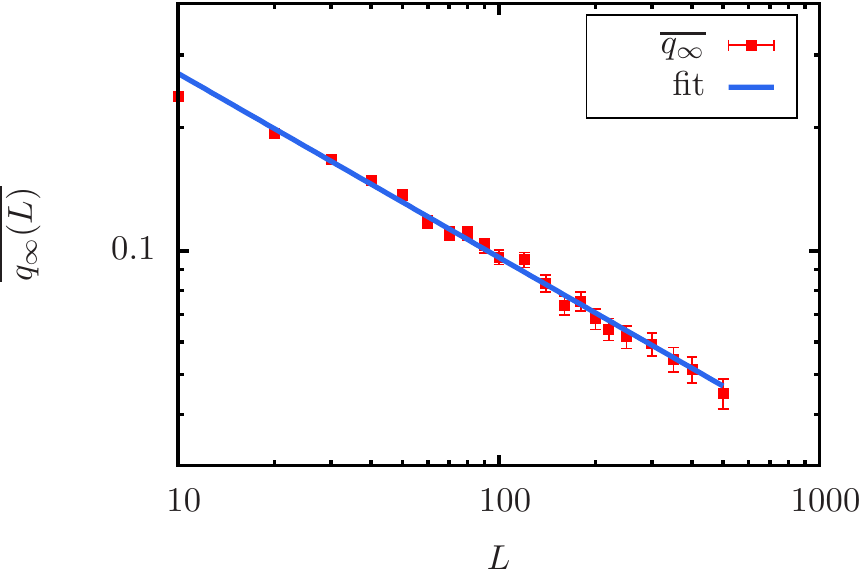}
\caption{$\qfinal$\ vs.\ $L$ for quenches to $T=0$. The solid line is the best power law fit
  for sizes 20 to 500, and corresponds to $\qfinal\sim L^{-0.46}$. From~\cite{YMNS13}.}
\label{fig:MSF}
\end{figure}

Next consider $\qfinal$, the finite size behavior of the absorbing
value of $\qlt$, shown in Fig.~\ref{fig:MSF}.  Ye~{\it et al.\/}~found
that the data were well fit by a power law of the form $\qfinal = a
L^{-b}$.  Their best estimate of $b$, taking into account both
statistical and possible systematic errors, was $0.46 \pm 0.02$.

Summarizing, the main results are:

\begin{enumerate}

\item Heritability exponent: $\qmean \sim d t^{-\cm}$ and $\cm=0.22 \pm 0.02$.

\item Size dependence: For finite $L$ and $T=0$, $\qfinal  \sim a L^{-b}$ with $b = 0.46 \pm 0.02$.

\item  Finite size scaling considerations suggest that $b=2\cm$, consistent
with the numerically determined values and with the exact $1D$ values.

\end{enumerate}

\section{Random ferromagnets and spin glasses in finite dimensions}
\label{sec:disordered}

In~\cite{YGMNS17}, the nature vs.~nurture question was studied in random Ising ferromagnets and Edwards-Anderson~(EA) spin glasses~\cite{EA75} in dimensions greater than one. Both use the Hamiltonian~(\ref{eq:disordered}), 
the difference being in the form of the probability distribution from which the couplings $J_{xy}$ are chosen. In both cases the $J_{xy}$ are i.i.d.~random variables, but in the ferromagnetic case the couplings are chosen from a continuous distribution supported on nonnegative real numbers, and in the spin glass the support of the coupling distribution lies on both positive and negative real numbers. Specifically, in~\cite{YGMNS17} the coupling distribution for the~EA spin glass in $d$~dimensions was taken to be a normal distribution with mean zero and variance one, and for the random ferromagnet the distribution was taken to be a one-sided Gaussian, in which each bond is chosen as the absolute value of a standard Gaussian random variable (again with mean zero and variance one).

In Sect.~\ref{sec:uniform}, it was observed that both of these models belong to class~${\cal F}$ in all dimensions; therefore, the proper quantity to study is the dynamical order parameter~$q_D$.  We showed in~Sect.~\ref{subsec:1D} that in any finite dimension, $q_D$ lies strictly between 0 and 1. We also know that the random ferromagnet and spin glass are identical (up to a trivial gauge transformation) in an infinite~$1D$ chain, and that $q_D=1/2$. As also discussed in Sect.~\ref{subsec:1D}, a  (poor) lower bound for $q_D$ is provided by the density of bully bonds $\rho_b(d)$, which goes to 0 as $d\to\infty$. A central question posed in~\cite{YGMNS17} is then: does $q_D(d)\to 0$ as dimension $d\to\infty$?

\begin{figure}[H]
\vskip 0.3cm
\centering  
\includegraphics[scale=1]{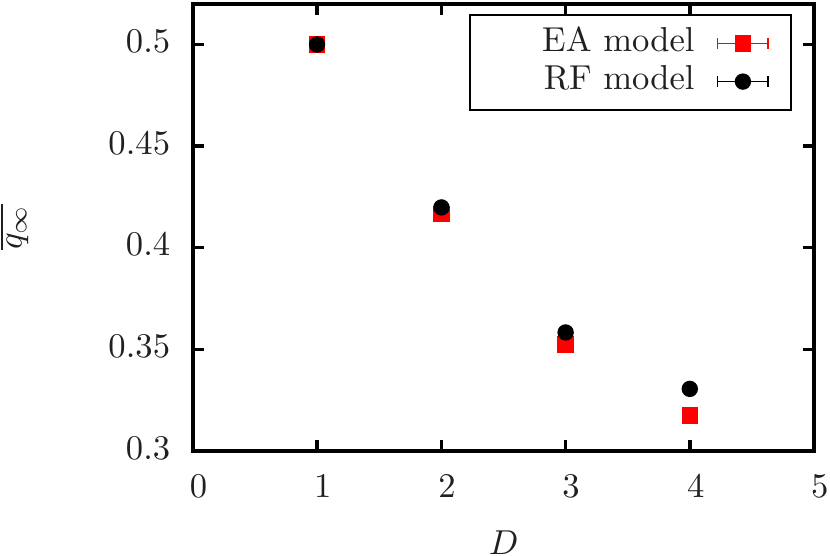}
\caption{The average twin overlap in the absorbing state for the largest systems size $\qfinalNmax$ vs.~dimension $d$ for the Edwards-Anderson spin glass (red squares) and random ferromagnet (black circles). From~\cite{YGMNS17}.} 
\label{fig:QD} 
\end{figure}

Fig.~\ref{fig:QD} shows numerically derived values for both systems in 2, 3, and 4 dimensions.

The results for the random ferromagnet are quite close to those of the Edwards-Anderson model, especially in $d=2$.   This suggests the possibility that frustration, present in the spin glass but not the random ferromagnet, plays little or no role in the nature vs.~nurture problem for low dimensionality.  This could change, however, as dimension increases; indeed, Fig.~\ref{fig:QD} indicates that as dimension increases the difference between $\qfinalNmax$ for the spin glass and random ferromagnet likewise increases, with $q_D$(random ferromagnet)$>q_D$(spin glass).  

While $q_D$ monotonically decreases with dimension up to 4, limits on what can be done numerically at this time preclude going to higher dimensions, and the question of the behavior of $q_D(d)$ as $d\to\infty$ remains open. Although the results from~\cite{YGMNS17} are consistent with the conjecture that $q_D(d)\to 0$ as $d\to\infty$, they don't preclude other possibilities. One might then take a look at the behavior of the random Curie-Weiss ferromagnet and the Sherrington-Kirkpatrick~(SK)~\cite{SK75} infinite-range spin glass to infer the high-dimensional behavior of the finite-dimensional random ferromagnet and spin glass. As we will see in the next section, however, the mean-field results are surprising, and so we will return to this question there.

\begin{figure}[H]
\centering  
\includegraphics[scale=0.5]{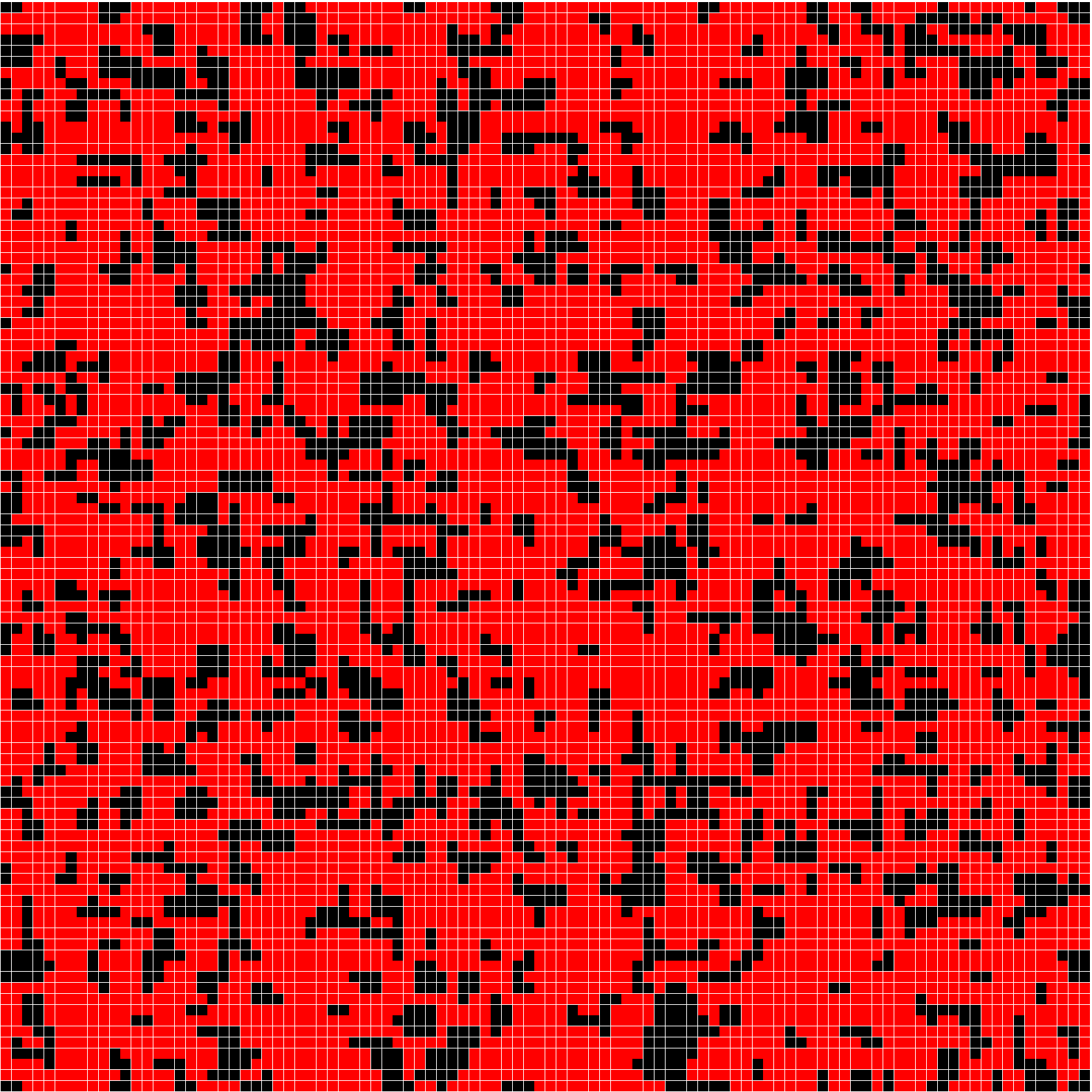}
\caption{(color online). The overlap configuration for a typical pair of final states for the two-dimensional Edwards-Anderson model with $L=100$.  Like spins percolate and are shown in red while unlike spins are shown in black.} 
\label{fig:config} 
\end{figure}

Before leaving this section, it is worth noting that several other features of the nature vs.~nurture problem besides $q_D$ were studied in~\cite{YGMNS17}. These include convergence rates of overlaps as the number of spins~$N$ increases and the mean survival time~$\tau(N)$ (i.e., the average number of spin flips per spin as a function of system size and dimension). The reader is referred to~\cite{YGMNS17} for a discussion of these quantities. One further study is especially interesting and will be briefly mentioned here, namely the spatial structure of the overlap in the final state.  It is natural to ask whether ``like spins'' (i.e., spins whose final state is the same) percolate in the twin samples. Fig.~\ref{fig:config} shows the overlap configuration for a typical pair of final states for the two-dimensional EA~spin glass with $L=100$.  From the figure it is indeed seen that like spins (shown in red) are well above the percolation threshold.  (Note that there are no isolated singleton unlike spins since those must be eliminated by the zero-temperature dynamics.)

\section{Mean-field models}
\label{section:mf}

We begin by examining the SK~model, which is the spin glass whose $N$ spins lie on the nodes of the complete graph. Its Hamiltonian is 
\begin{equation}
\label{eq:SK}
{\cal H}=-\frac{1}{\sqrt{N}}\sum\limits_{i<j}J_{ij}\sigma_i\sigma_j\, .
\end{equation}
The couplings are again chosen from a normal distribution with mean zero and 
variance one, and the rescaling factor
$N^{-1/2}$ ensures a sensible thermodynamic limit of the energy and free energy per spin. 

\begin{figure}[H]
\centering 
\includegraphics[scale=1]{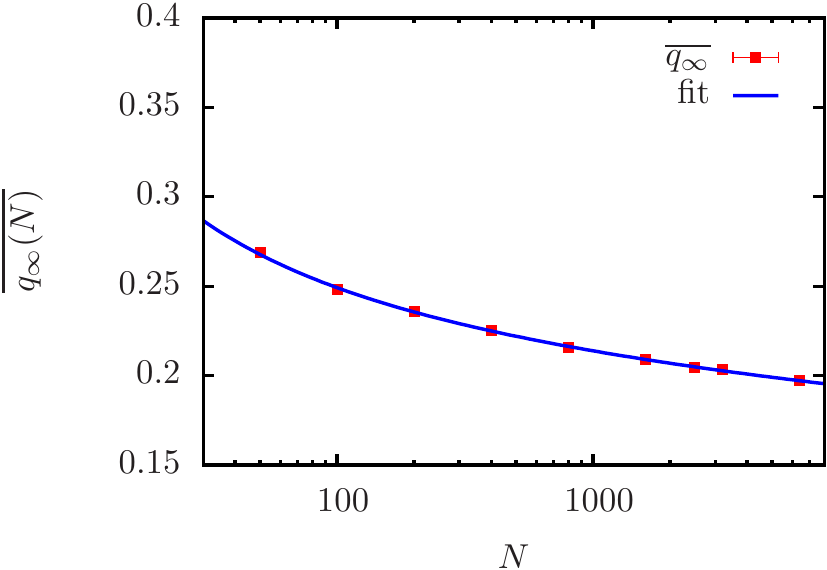}
\caption{Simulation results for $\qfinalN$ vs.!$N$ for the SK model.  The curve is the highest quality fit:  $\qfinalN = \displaystyle\frac{a}{(\log{N})^{1/3}}+ \frac{b}{N}$, with $a$ and $b$ both close to~0.4. From~\cite{YGMNS17}.}
\label{fig:FIT} 
\end{figure}

Figure \ref{fig:FIT} is a plot of $\qfinalN$  
as a function of $N$.  While it is clear that $\qfinalN$ is decreasing with $N$, it again is not obvious whether $\qfinalinfN$ is zero or greater than zero.  

The best fit implies that $\qfinalinfN\to 0$, although slowly. However, a different fit --- not as good, but still reasonable --- implies a nonzero limit. For details, see~\cite{YGMNS17}. Nevertheless, a heuristic argument presented in~\cite{YGMNS17} suggests that it is most reasonable to expect that indeed $\qfinalinfN\to 0$. The main idea is the following: it is known that the SK model has exponentially many (in $N$) one-spin-flip metastable states~\cite{BM80}, and moreover numerics indicate that the number of spin flips grows linearly with $N$~\cite{YGMNS17}. Given the $O(N)$ distance traveled by the SK model on the state space hypercube to find an absorbing state, and given that these states have been shown to be uncorrelated~\cite{BM80}, the decay of $q_\infty\to 0$ as $N\to\infty$ appears to be its most likely behavior.

Again, it is worth noting that several other dynamical behaviors of the SK~model were studied, including the median time for a system of $N$~spins to reach the absorbing state, the fraction of active spins (those that have not finished flipping) as a function of time and system size, and the energy per spin as a function of time. The reader is referred to~\cite{YGMNS17} for a discussion of these behaviors.

The evidence for $\qfinalinfN\to 0$ as $N\to\infty$ for the SK model may be taken to imply that $q_D(d)\to 0$ for the EA~spin glass as $d\to\infty$. This is reasonable, and even likely to be correct, but a cautionary note should be added: the behavior of the Curie-Weiss random ferromagnet is completely different --- in fact, for this model $\qfinalinfN\to 1$ as $N\to\infty$~\cite{YGMNS17}! While presumably the information in the initial state is completely lost in the SK model for large systems at long times, for the random ferromagnet it's completely retained. The profound difference between nature and nurture for these models demonstrates that frustration plays a centrally important role in infinite dimensions, and so it may play an important role in high but finite dimensions as well.

It is easy to understand why the Curie-Weiss model behaves this way. Consider first the uniform case, where all couplings have equal magnitude. A typical initial condition will have an excess 
(of order $\sqrt{N}$) of spins in one state (say the plus state) over the other, so every spin feels the same positive internal field, and this can only increase with time. Given the usual Glauber dynamics, it's clear that the final state will then be all plus, so the initial condition completely determines the final configuration. The only initial conditions in which this will not be true is that for which $-1\le\sum_{i=1}^N\sigma_i\le 1$. But the contribution to $q_\infty$ from such configurations goes to zero (as $N^{-1/2}$) as $N\to \infty$. 

One expects exactly the same behavior for the random Curie-Weiss model, but the proof is considerably more difficult. As before, a typical initial spin configuration will have $O(\sqrt N)$ excess of plus or minus spins, but now the internal fields acting on the spins will vary, and if the initial excess of spins is positive, the internal fields on some spins could be negative. One therefore has to study the {\it distribution\/} of the internal fields at each site, which at time zero has positive mean. The main technical issue is to show that the fraction of sites with positive internal field increases steadily with time.

A proof covering the case of the randomly diluted Curie-Weiss ferromagnet, where the $J_{ij}$'s are chosen from Ber$(p)$ for some fixed $p\in(0,1)$, showing that again  $\qfinalinfN\to 1$ as $N\to\infty$, will appear soon~\cite{GNS17}.  The proof demonstrates that after a time of order $N^{1/2+\epsilon}$, where $\epsilon > 0$ is independent of $N$, every site has positive internal field with probability going to one as $N\to\infty$. At this point, the dynamics monotonically leads to absorption into the all-plus state, so that $q_\infty\to 1$ as $N\to\infty$.

The proof further demonstrates that the final state of the system is one of the two uniform states. One possibility is then that the random Curie-Weiss ferromagnet possesses many metastable states (like finite-dimensional random ferromagnets and spin glasses~\cite{NS99a} or the SK~model~\cite{BM80}), but that the dynamics somehow avoids them. More likely, though, is the possibility that the random Curie-Weiss model possesses no metastable states (in the sense that their number falls to zero as $N\to\infty$), or else too few for the system to find. Recent numerical work by Wang~{\it et al.\/}~\cite{WGNS17} strongly suggests that the latter is what occurs.

These results lead to a problem in interpreting the finite-dimensional random ferromagnetic model: either $q_D(d)$ reaches a minimum at some finite $d$ and then increases to 1 as $d\to\infty$, or else $q_D(d)$ falls to zero as conjectured and the infinite-dimensional limit is singular for the nature vs.~nurture problem in the random ferromagnet as $d\to\infty$. (There are other possibilities, of course, but these are the most plausible alternatives.) In~\cite{YGMNS17} a heuristic argument is presented in support of the latter alternative, but the problem remains open in the absence of a more detailed argument (or preferably, a proof).

Other models were studied as well in~\cite{YGMNS17}, in particular the random energy model (REM) of Derrida~\cite{Derrida80,Derrida81}, where it was again found that $q_\infty\to 1$. The argument can again be found in~\cite{YGMNS17} and will not be repeated here. What is important to note is that all of these studies indicate strongly that two conditions appear to be necessary (though perhaps not sufficient) in order for $q_\infty < 1$. The first is the presence of a large number of uncorrelated metastable states, so that the system has many possible final states whose overlap is small. This condition is satisfied for the EA model and random ferromagnet in all finite dimensions (although the metastable states retain some correlations, which presumably decrease to zero as $d\to\infty$), as well as by the SK model and the REM.  The second, equally important, condition is that a finite system undergoes at least $O(N)$ spin flips before reaching the absorbing state, and correspondingly for an infinite system, the average number of spin flips per site is strictly positive. This is the case for the EA model and random ferromagnet in all finite dimensions, as well as for the SK model and both mean-field ferromagnets. Of the mean-field models studied, only the SK model satisfies both conditions.

In contrast, the random Curie-Weiss ferromagnet and REM have $q_\infty\to 1$, but for very different reasons. In the random Curie-Weiss ferromagnet, there are $O(N)$ steps in the random walk on the configuration space hypercube (with $2^N$ vertices) but (presumably) no metastable states to trap the walk before reaching the absorbing uniform final state consistent with the initial configuration. The REM, on the other hand, has many metastable states, but its random walk travels only $O(\log N)$ steps before being trapped in a metastable state~\cite{YGMNS17,KL87}; hence the overlap with the initial state approaches~1 as $N\to\infty$.

\section{Conclusion}
\label{sec:conclusion}

This relatively brief review has touched on some of the central topics in the nature vs.~nurture problem, but omitted many interesting issues and results which can be further pursued in the papers listed in the bibliography. There are numerous outstanding questions and open problems, but perhaps the most fundamental ones are the following.

\begin{enumerate}

\item Determine the dynamical class (${\cal I}$, ${\cal F}$, or ${\cal M}$) of the uniform Ising ferromagnet in $d\ge 3$. For those dimensions belonging to class ${\cal I}$ (or ${\cal M}$), determine the heritability exponent and study its behavior as a function of dimension. How does it relate to the persistence exponent, particularly as dimension increases? 

\item What is the behavior of $q_D(d)$ for the random ferromagnet and EA~spin glass as a function of dimension? Does $q_D(d)\to 0$ as $d\to\infty$? The case of the random ferromagnet is particularly interesting: does it begin to increase at some finite $d$, or is the dynamical behavior singular at $d=\infty$? Answering this question would shed light on the equally important question of the role of frustration in finite dimensions: is it as important as it appears to be in infinite dimensions?

\item Prove (or disprove) that  $\qfinalinfN\to 0$ as $N\to\infty$ for the SK model.

\item A particularly difficult problem is disordered systems with {\it discrete\/} coupling distributions, most notably the $\pm J$~spin glass. This was shown in~\cite{GNS00} to be in class~${\cal M}$ in two dimensions. What is its dynamical behavior in higher dimensions? The nature vs.~nurture question has not been studied at all in this model, nor more generally has a serious attempt been made to understand how to approach this problem in the context of class-${\cal M}$ models.

\end{enumerate}

If nothing else, a major aim of this review is to provide the reader with the sense that the nature vs.~nurture approach to dynamics constitutes a set of deep problems and rich phenomena whose explication can provide significant illumination on the dynamical behavior of both ordered and disordered statistical mechanical systems.

\begin{acknowledgements} 
I thank my collaborators on various aspects of this work --- Reza Gheissari, Jon Machta, Chuck Newman, Paolo Maurilo de Oliveira, Vladas Sidoravicius, Lily Wang, and Jing Ye --- for a productive and enjoyable collaboration. And of course, happy birthday (again) to Chuck!
\end{acknowledgements}

\small\def\em{\it} \newcommand{\noopsort}[1]{} \newcommand{\printfirst}[2]{#1}
  \newcommand{\singleletter}[1]{#1} \newcommand{\switchargs}[2]{#2#1}

\end{document}